\documentclass[preprint,
preprintnumbers,amsmath,amssymb,superscriptaddress]{revtex4-2}

\usepackage{fullpage,amsthm,amsmath,amsfonts,bm,times,color,bbm}
\usepackage{graphicx}
\usepackage[english]{babel}
\usepackage{mathtools,color}
\usepackage{ccaption}
\usepackage{graphicx}
\usepackage{array}
\usepackage{float}
\usepackage{wrapfig}
\usepackage{caption}
\usepackage{subcaption}
\usepackage{hyperref}
\usepackage{romannum}
\usepackage{amsmath}
\usepackage{amssymb}
\usepackage{multirow}
\usepackage{booktabs}
\usepackage{makecell}
\usepackage{lineno}
\usepackage[flushleft]{threeparttable}

\usepackage{soul} 

\makeatother

\makeatletter

\begin{document}
	
	
	\pagenumbering{arabic}
	\title{Regional Greening as a `Positive' Tipping Phenomenon}
	\author{Yu Sun}
	\affiliation{School of Systems Science/Institute of Nonequilibrium Systems, Beijing Normal University,  Beijing 100875, China}	
 
	\author{Teng Liu}
	\affiliation{School of Systems Science/Institute of Nonequilibrium Systems, Beijing Normal University,  Beijing 100875, China}	
    
        \author{Shang Wang}
	\affiliation{School of Systems Science/Institute of Nonequilibrium Systems, Beijing Normal University,  Beijing 100875, China}	
\author{Jun Meng}
\affiliation{School of Science, Beijing University of Posts and Telecommunications, Beijing 100876, China}
\affiliation{Potsdam Institute for Climate Impact Research, Potsdam 14412, Germany}
\author{Yongwen Zhang}
\affiliation{Data Science Research Center, Faculty of Science, Kunming University of Science and Technology,
Kunming 650500, China}
\author{Saini Yang}
\affiliation{State Key Laboratory of Earth Surface Processes and Resource Ecology, Beijing Normal University, Beijing 100875, China}
\affiliation{School of National Safety and Emergency Management, Beijing Normal University, Beijing 100875, China}
\author{Xiaosong Chen}
\affiliation{School of Systems Science/Institute of Nonequilibrium Systems, Beijing Normal University,  Beijing 100875, China}
        \author{Deliang Chen}
        \affiliation{Department of Earth Sciences, University of Gothenburg, Gothenburg 40530, Sweden}

        \author{J\"urgen Kurths}
        \affiliation{Potsdam Institute for Climate Impact Research, Potsdam 14412, Germany}
        \affiliation{Department of Physics, Humboldt University, Berlin 10099, Germany}
\author{Shlomo Havlin}
\affiliation{Department of Physics, Bar Ilan University, Ramat Gan 52900, Israel}
        \author{Hans Joachim Schellnhuber}
        \affiliation{Potsdam Institute for Climate Impact Research, Potsdam 14412, Germany}

        \author{Jingfang Fan}%
	\email{jingfang@bnu.edu.cn}
	\affiliation{School of Systems Science/Institute of Nonequilibrium Systems, Beijing Normal University,  Beijing 100875, China}
	\affiliation{Potsdam Institute for Climate Impact Research, Potsdam 14412, Germany}
	
	\begin{abstract}
Earth system tipping elements have been predominantly investigated for their potential to trigger \textit{negative} ecological, climatic, and societal shifts. 
Yet, an overlooked but seminal avenue exists in the form of \textit{positive} tipping phenomena, whose underlying mechanisms and benefits remain largely underexplored. To bridge this gap, our research introduces a fundamental percolation-based framework to assess the criticality and resilience of planetary terrestrial vegetation systems. Leveraging high-resolution satellite data, we focus on greening-induced positive tipping dynamics driven by global warming. We feature the Qinghai-Tibetan Plateau (QTP) and the Sahel region as contrasting yet analogous case studies.  Our analysis uncovers an intriguing phenomenon where vegetation fragmentation aligns with a percolation threshold, exhibiting a scale-invariant pattern characterized by nearly perfect power laws with three critical exponents. Remarkably, contrary to conventional destructive tipping elements, these regions act as favorable tipping elements, transitioning from fragmented to cohesive vegetation patterns due to anthropogenic climate change and afforestation efforts. Furthermore, we propose an \textit{optimal resilience enhancement model} to reinforce vegetation robustness while minimizing socio-economic costs. This study provides valuable insights into the favorable aspects of tipping elements under climate change and offers effective strategies for enhancing ecological resilience against environmental threats.

	\end{abstract}
	\date{\today}
	\maketitle

\section{Main}

Tipping elements, integral components of the Earth System, possess the capacity for sudden and irreversible state shifts when specific thresholds are exceeded \cite{Lenton2008,lenton2019Nature,armstrong_mckay_exceeding_2022}. These tipping points can precipitate catastrophic cascading failures across ecological, climatic, and societal domains, often in response to minor perturbations. Such tipping points are not restricted to Earth System but manifest in a variety of other complex systems, including financial markets \cite{May2008Nature,Kambhu2007EconomicPolicyReview}, ecosystems \cite{Hirota2011Science, Scheffer2001,Kefi2007,Drake2010,Lever2014,Jiang2018}, and social dynamics \cite{Centola2018,Otto2020}. While the majority of research has focused on the negative aspects of tipping points—namely, their potential to induce collapses or disrupt ecosystems—certain tipping elements can also drive favorable outcomes. These positive tipping elements offer a largely untapped avenue for mitigating the adverse impacts of climate change by fostering resilience and adaptive capacity across diverse domains. In this paper, we introduce a novel percolation-based framework to assess the criticality and resilience of planetary vegetation systems. We focus on two contrasting yet analogous regions—the Qinghai-Tibetan Plateau (QTP) and the Sahel—which serve as case studies for examining the complex interplay of tipping elements within the broader context of climate change and human activities.

\begin{figure}[hbpt]
          \centering
          \includegraphics[width=0.85\linewidth]{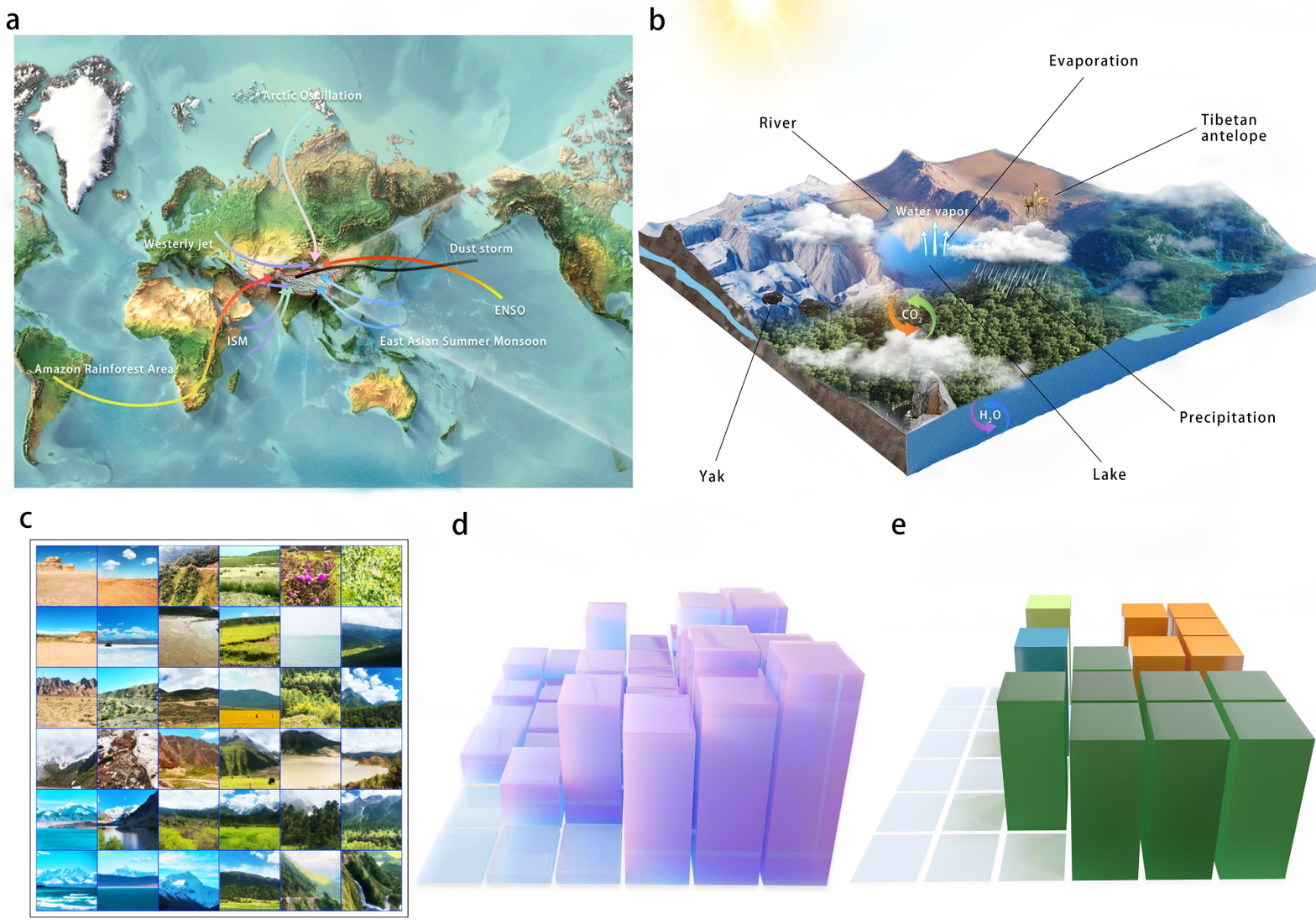}
          \caption{ { \bf A schematic representation of the complex conditions within the Qinghai-Tibetan Plateau (QTP) and an illustration of the percolation-based framework.}
          {\bf a}, The QTP, as a key component of Earth’s climate system, dynamically interacts with various climatic phenomena. Arrows in different colours, along with descriptions, indicate interactions with the westerly jet, Indian Summer Monsoon (ISM), East Asian Summer Monsoon, Arctic Oscillation, El Ni\~{n}o-Southern Oscillation (ENSO), dust storms, and the Amazon rainforest area.
          {\bf b}, A schematic representation of the QTP's biospheric, cryospheric, and hydrological conditions, highlights the intricate interplay between these elements.
          {\bf c}, Photos captured during June-July of 2021 and 2022 ($\copyright$ 2023 Jingfang Fan. All rights reserved) illustrate  the diverse vegetation conditions observed in the QTP, providing visual examples of the area's vegetation.
          {\bf d}, 
          The corresponding Enhanced Vegetation Index (EVI) values to various vegetation conditions shown in {\bf c} are represented by the heights of the bars. Taller bars indicate higher EVI values, indicating denser vegetation.
          {\bf e}, The spatial distribution of fragments derived from the percolation-based framework, generated from the EVI configuration shown in Panel {\bf d},  is displayed. The gray background represents the nodes that were removed (EVI$<$EVI$_c$), whereas coloured bars represent fragments. Nodes belonging to the same fragment share the same colour, visually demarcating clusters of vegetation.}
          \label{fig1}%
        \end{figure}

The \textbf{QTP}, often referred to as the  ``Third Pole", spans an impressive 2.5 million km$^2$ with an average elevation of 4.5 km. Its massive glacier storage, second only to the Antarctic and Arctic, feeds Asia's ten largest rivers, serving nearly 800 million people downstream \cite{Yao2012NatureClimateChange, Immerzeel2019Nature, Yao2019BulletinoftheAmericanMeteorologicalSociety,Xu2008GeophysResLett}. The Plateau's unique topography influences various  atmospheric, cryospheric, hydrological, and biospheric  processes \cite{Yao2019BulletinoftheAmericanMeteorologicalSociety, Xu2008GeophysResLett,Yao2012NatureClimateChange,Yang2011ClimaticChange,Shen2015ProceedingsoftheNationalAcademyofSciencesoftheUnitedStatesofAmerica,Yao2012EnvironmentalDevelopment}, leading to broad climate patterns and establishes large-scale teleconnections with various climatic phenomena and tipping elements, such as the Indian Summer Monsoon  \cite{Cai2012EarthandPlanetaryScienceLetters,Sato2007MonthlyWeatherReview, Sills2010Science}, East Asian Summer Monsoon \cite{Yang2014GlobalandPlanetaryChange,Sills2010Science},  westerly jet \cite{Duan2012AdvancesinAtmosphericSciences}, Amazon rainforest  \cite{Liu2023NatureClimateChange},  Arctic Oscillation  \cite{Bi2022ClimateDynamics},  El Ni\~{n}o-Southern Oscillation \cite{Shaman2005JournalofClimate,Wu2012JournalofClimate},  dust storms  \cite{Huang2007GeophysResLett}, and others (Fig. \ref{fig1}a).
Alarmingly, it has experienced a rate of warming twice the global average over the past five decades  \cite{Chen2015ChineseScienceBulletin}, revealing its significant sensitivity to climate change.
Terrestrial vegetation on the QTP, primarily alpine steppe and meadow, plays a vital role in both local and global systemic stability. Climate change has a profound impact on QTP vegetation, affecting processes such as photosynthesis \cite{Zhu2016NatureClimateChange, Zhang2013ProceedingsoftheNationalAcademyofSciences, Shen2014AgriculturalandForestMeteorology} and tree-line patterns \cite{Liang2016ProceedingsoftheNationalAcademyofSciences, Gou2012Dendrochronologia}. This leads to feedback mechanisms including evaporative cooling \cite{Shen2015ProceedingsoftheNationalAcademyofSciencesoftheUnitedStatesofAmerica}, carbon sequestration \cite{Chen2020Geoderma}, and complex, self-reinforcing interconnections involving vegetation, air, ice, water, wildlife, and human activity (Fig. \ref{fig1}b).

In contrast, the \textbf{Sahel} region acts as Africa's climatic and ecological pivot, lying between the arid Sahara and the humid forests of West and Central Africa \cite{Lee_PG_2015}. Its northern fringe is defined by the Sahara desert, while its southern edge reveals a  sharp gradient in vegetation, demarcating the shift towards the monsoonal forests characteristic of West and Central Africa. Its unique position makes it particularly vulnerable to the impacts of climate change and anthropogenic activities, a fact extensively supported by scholarly research \cite{begueCan25yearTrend2011, lehouerouRangelandsSahel1980, Diffenbaugh_CC_2012}. Recent fluctuations in precipitation patterns have exerted discernible effects on both ecological systems and human societies  \cite{bathiany_co2-induced_2014, ouedraogo_re_greening_2014}. Notably, the Sahara has undergone periodic wet phases throughout the Quaternary period, with a proliferation of vegetation \cite{armstrong_north_2023}. Future projections suggest that climate change could further amplify this trend, resulting in even greater rainfall in the region \cite{pausata_greening_2020}.
Despite these insights, pinpointing the specific causative factors underlying these changes remains a complex analytical challenge  \cite{Nicholson_JoAE_2005, brovkin_stability_1998, vamborg_background_2014, xue_sahelian_2004}.


From a theoretical standpoint, understanding tipping points is crucial for predictive and adaptive strategies. Early-warning signals like lag-1 autoregression, variance, skewness, and detrended fluctuation analysis often rely on the principle of critical slowing down (CSD), which refers to the system's progressively slower recovery from perturbations as it approaches a tipping point  \cite{Scheffer2009Nature,Scheffer2012Science,livina_modified_2007}. Yet, the capacity of CSD-based analyses to capture the geometric complexity and emerging functional structures that stem from spatiotemporal variability within the system might be limited. Additionally, the accuracy and effectiveness of such indicators may be hindered by inadequate empirical observations over time.

To overcome the limitations of conventional early-warning signals, our percolation-based framework  employs high-resolution satellite data to evaluate the self-organized criticality (SOC) \cite{Bak1987PhysRevLett} and resilience of the vegetation in both the QTP and the Sahel region. Drawing a parallel to occupancy probability in percolation theory, each grid point represents different vegetation types, characterized by an associated Enhanced Vegetation Index (EVI) value indicating the degree of \textit{greenness}. Higher EVI values correspond to more abundant and thriving vegetation, similar to higher occupancy probability signifies a greater likelihood of a site being occupied in a percolation model. In our framework, when neighboring cells have EVI values surpassing a predefined threshold (EVI$_c$), they form a vegetation fragment (or cluster) \cite{Taubert2018Nature}. This process is illustrated in Fig. \ref{fig1}c-e. We present photos taken during  June-July of 2021 and 2022, displayed in Fig. \ref{fig1}c,  showcasing the diversity of vegetation present in the QTP. The corresponding EVI values are depicted as bar heights in Fig. \ref{fig1}d and values in Fig. S1a. Additionally, after removing nodes with EVI values below EVI$_c$, the configuration of vegetation fragments can be observed in Fig. \ref{fig1}e and Fig. S1b. 

This framework delivers a unique spatial perspective, enabling the identification of functional fragmentation structures – the disintegration of ecosystems into isolated units – and the detection of geometric tipping properties like scale invariance and self-similarity in vegetation patterns. Notably, our calculated percolation threshold EVI$_c$ approaches an empirically established ecological tipping point of approximately 0.2, delineating the accepted threshold between sparse or nonexistent vegetation and regions of healthy vegetation over the past two decades. Concurrently, the accompanying power law distributions also support the approach to this tipping point, exhibiting three identical exponents for fragment size distribution ($\tau$), mass fractal dimension ($D_f$), and external diameter fractal dimension ($D_e$). Through this method, critical areas requiring protection can be identified, facilitating the development of effective strategies to tackle the challenges posed by factors such as climate change.  This insight allows policymakers and conservationists to prioritize their efforts and allocate resources more efficiently by pinpointing regions that are particularly vulnerable or essential for ecosystem health.

\section{Results}
To conduct our percolation analysis \cite{bunde2012fractals}, EVI \cite{Huete2002RemoteSensingofEnvironment}, a well-established index of greenness and more effective than other vegetation indices, such as the Normalized Difference Vegetation Index, is used. Our analysis utilizes high spatial resolution QTP EVI maps (25.99°N-39.82°N, 73.49°E-104.63°E), which contain approximately 85.5 million pixels, and Sahel EVI maps (9.45°N-20.07°N, 17.40°W-38.29°E), which contain approximately 117.3 million pixels. Each pixel measures 250m $\times$ 250m. 
Our study focuses on the QTP and Sahel region during the summer months (July-September) of the Northern Hemisphere. This period coincides with the peak growing season when vegetation cover is at its highest within these area. 
Please refer to the Data Section (\ref{datasection}) for comprehensive details and references regarding the data used in this study.

\subsection{Criticality of vegetation in the Qinghai-Tibetan Plateau and Sahel}
To assess the criticality of vegetation in the QTP, we perform a site percolation analysis using the average EVI image from the recent five-year period (2017-2021). 
The spatial pattern of the EVI is depicted  in Fig. \ref{fig2}a.
In our analysis, we're systematically attacking and removing nodes (i.e., pixels) with EVI values that fall below a given threshold. The resulting fragments are tracked at each step using the Newman-Ziff algorithm \cite{Newman2000PhysRevLett}. Fig. \ref{fig2}b illustrates the relative sizes of the largest fragment, $G_1$, and the second largest fragment, $G_2$, as a function of EVI.
We observe a sudden and dramatic jump in the size of the largest fragment ($G_1$) at a percolation threshold (EVI$_c$), which is approximately 0.2 (indicated by the vertical black dashed line). Concurrently, 
the maximum size of the second-largest fragment ($G_2$) also peaks at this threshold.
Fig. \ref{fig2}c displays the greenness fragmentation structures at EVI$_c$, highlighting only fragments with a size (node number) larger than 100. The structure reveals two major fragments, with a critical node at an EVI$_c$ value located near (36.07$^\circ N$, 101.96$^\circ E$) acting as a connection between these fragments. 
In percolation theory, the percolation threshold is determined by $G_1$ \cite{fan_universal_2020} and $G_2$ \cite{margolina_size_1982}. Ecologically, an EVI value of 0.2 typically represents the threshold between unhealthy (without or sparse vegetation cover) and healthy vegetation \cite{liu_feedback_1995}. The strong agreement between our theoretical percolation threshold EVI$_c$ and empirical values highlights the criticality of vegetation in the QTP. 

Next, we analyse the fragment size distribution, $n_s$, at EVI$_c$, as described in Section (\ref{clustersizefitting}). Our analysis reveals a total of over 87,000 fragments distributed throughout the QTP, with sizes spanning seven orders of magnitude. Intriguingly, the fragment size distribution follows a perfect power-law relationship: $n_s \sim s^{-\tau}$, where the exponent $\tau$ is approximately 2.04, as depicted in Fig. \ref{fig2}d. This empirical exponent aligns closely with the theoretical value of $\tau = 187/91 \approx 2.05$ as predicted by the classical percolation theory for a 2D lattice \cite{bunde2012fractals,stauffer2018introduction}.

Furthermore, we investigate the fractal properties of the giant fragment at EVI$_c$. Using the \textit{box renormalization method} (refer to Methods) and examining the mass and external perimeter of the giant fragment at various image resolutions, we confirm that both the inner structure and external boundary of the giant fragment at EVI$_c$ exhibit self-similarity (see Section \ref{fractaldimension}). Our analysis, shown in Fig. \ref{fig2}e, yields empirical fractal dimensions for the mass ($D_f\approx 1.93$) and external perimeter ($D_e \approx 1.40$). Remarkably, these empirical exponents, $D_f$ and $D_e$ are consistent with the theoretical values of $D_f = 91/48\approx 1.90$ and $D_e = 4/3\approx 1.33$ at the percolation threshold for a 2D lattice (refer to Table. \ref{tab1}).

To provide a comparison, we also introduce a null model generated by shuffling the original EVI data (100 independent realizations), which represents the classical uncorrelated site percolation. The results, depicted in Fig. \ref{fig2}d, Fig. S2a, and Fig. S2b, demonstrate consistent power-law characteristics in the fragment size distribution and fractal substructures between the empirical and null models, with similar exponents $\tau$, $D_f$, and $D_e$ (Table. \ref{tab1}) for the QTP. These findings strongly suggest that the greenness structures of the QTP at EVI$_c$ are in a critical state,  indicative of a system operating at the SOC stage.

        \begin{table}[b]
            \begin{center}
            \caption{{\bf Scaling fragmentation of vegetation cover on the Qinghai-Tibetan Plateau (QTP) and Sahel. 
            }
            Critical exponents $\tau$, $D_f$, and $D_e$ are calculated in the empirical and null models, as well as theoretical exponents corresponding to 2D classical percolation theory at the percolation threshold. At the critical point, the exponent of the fragment size distribution $\tau$ is further related to the fractal dimension $D_f$ via the hyperscaling relationship 
            $\tau = 1+ d/{D_f}$ \cite{bunde2012fractals,stauffer2018introduction,ben-avraham_diffusion_2005}. 
            }\label{tab1}%
            \begin{tabular}{|m{6.5cm}| m{2.4cm} | m{2.4cm}| m{2.5cm} | m{2.4cm}|} 
            \toprule
             &QTP&  Sahel & Null model & Theory  \\
            \botrule
            Fragments size distribution exponent  $\tau$ & 2.04 ($\pm 0.02$)& 2.06  ($\pm 0.02$) & 2.06 ($\pm 0.0006$) &  187/91 \cite{stauffer2018introduction}     \\
            Mass fractal dimension $D_f$  & 1.93 ($\pm 0.07$)  & 1.90  ($\pm 0.06$)  & 1.91 ($\pm 0.02$)  &  91/48 \cite{nijs_relation_1979}     \\
            External diameter fractal dimension $D_e$  & 1.40 ($\pm 0.04$)  & 1.19  ($\pm 0.03$)  & 1.33 ($\pm 0.01$) &  4/3 \cite{Grossman1987JournalofPhysicsAMathematicalandGeneral}    \\
            \toprule
            \end{tabular}
            \end{center}
        \end{table}

Building on our analytical percolation framework, we extend our assessment to evaluate the Sahel region's ecological criticality,  conducting site percolation analyses for the period 2013-2017. 
The spatial distribution of the EVI values across the Sahel is represented in Fig. \ref{fig2}f. In a manner analogous to our work on the QTP, Fig. \ref{fig2}g presents the relative sizes of the largest fragment ($G_1$) and the second-largest fragment ($G_2$), respectively. These are charted against varying EVI values, allowing us to identify the percolation threshold (EVI$_c$) at approximately 0.2, as indicated by the vertical black dashed line.
Fig. \ref{fig2}h illustrates the Sahel's greenness fragmentation structures at this critical EVI$_c$ level. Specifically, it highlights two major vegetation fragments centered around the critical node near (14.92$^\circ$N, 3.86$^\circ$W). 
The distribution of fragment sizes at EVI$_c$ also follows a power-law formula $n_s \sim s^{-\tau}$, where the exponent $\tau$ is calculated to be approximately 2.06, as demonstrated in Fig. \ref{fig2}i.
Empirical fractal dimensions for mass ($D_f$ $\approx$ 1.90) and external perimeter ($D_e$ $\approx$ 1.19) are determined in Fig. 2j.
Notably, the three empirical exponents $\tau$, $D_f$, and $D_e$ for the Sahel are almost consistent with the theoretical values established for a percolation in a 2D lattice (refer to Table. \ref{tab1}).

\subsection{Greening-induced tipping point}

To investigate the spatio-temporal patterns of vegetation on the QTP and Sahel, we employ percolation procedures on a series (2001-2021) of temporally evolving EVI images, each representing the temporal average of pixel-by-pixel EVI values within a 5-year sliding window. The sliding window moves forward by 2 years, resulting in adjacent windows with a temporal overlap of 3 years.
We analyse the percolation phase diagram and determine the EVI$_c$ for each sliding window, as illustrated in Fig. S3. In analogy to percolation theory, the resilience state of QTP's vegetation can be evaluated through  EVI$_c$. This threshold indicates the critical point at which the cohesive connectivity within the vegetation starts to break down \cite{cohen_resilience_2000}. A higher EVI$_c$ denotes a better state that more extensive vegetation damage is required to disrupt the ecosystem. As depicted in Fig. \ref{fig3}b, we observe an overall trend of EVI$_c$ increasing from 0.17 to 0.23 over the past two decades, indicating a shift toward a positive tipping point and enhanced ecosystem resilience. To further assess the changes in vegetation patterns, we compare the coverage area of the largest fragment in the QTP between the periods of 2001-2005 and 2017-2021 at the same threshold EVI$_c$=0.23 (before the jump). As shown in Fig. S4, we observe an increase of approximately \textbf{16.2}\% in the coverage area percentage of the QTP's largest fragment, accompanied by greenness in the northeast and the clustering of numerous small greenness fragments.  The enhancement in greenness resilience suggests that the QTP has been transitioning from a fragmented to a cohesive phase, acting as a \textit{positive tipping element}.

Additionally, we verify the power-law form of the fragment size distribution for each sliding window, as shown in Fig. S5. Remarkably, we observe a significant trend in the empirical fragment size distribution exponent $\tau$ over the period of 2001-2021 for the QTP. As depicted in Fig. \ref{fig3}d, $\tau$ progressively approaches the theoretical value of  $\tau \approx 2.05$. This trend suggests increasing alignment of the fragmentation of QTP's vegetation cover with classical percolation theory. Furthermore, we analyse the goodness of power-law fits for the fractal dimensions $D_f$ and $D_e$ (shown in Fig. S6 and Fig. S7). The results reveal that both fractal dimensions also exhibit a significant trend toward the theoretical values, $D_f =1.90$ and $D_e = 1.33$, as shown in Fig. S8. These findings suggest that since approximately 2010, the greenness of the QTP has been converging towards the tipping point, supported by the proximity of the critical exponents to its theoretical value.

Both observational and modelling studies have demonstrated that recent climate change has impacted the structure and ecological functioning of QTP vegetation \cite{Shen2015ProceedingsoftheNationalAcademyofSciencesoftheUnitedStatesofAmerica,Chen2013GlobChangeBiol,Tan2010GlobalBiogeochemCycles,Wang2012Ecology}. In marked contrast we find here, in Fig. \ref{fig3}a, the spatial pattern of significant EVI change trends from 2001 to 2021 reveals a general positively greening trend, particularly in the Northeast, Northwest, and Southwest regions of the QTP. Fig. S9 reveals distinct temporal fluctuations in the EVI values among the critical nodes identified within the QTP.

To investigate the potential climatic drivers behind the observed criticality and enhanced resilience of the QTP, we examine the relationship between the evolution of EVI and temperature as well as precipitation patterns across the region. Our investigation reveals a similar trend in both, the EVI$_c$ and the warming of the QTP (Pearson correlation coefficient $r$ = 0.88,  statistically significant at $p<10^{-4}$), as demonstrated by comparing EVI with the average temperature from June to August (Fig. \ref{fig3}c). Additionally, increased precipitation was identified as a key factor contributing to the enhancement of greenness. The variation in vegetation intensity aligns with the variation in precipitation, exhibiting a delay of approximately two years (Fig. S10). Pixel-by-pixel cross-correlation analysis with time lags reveals the relationship between the EVI series and precipitation series, identifying precipitation-induced spatial patterns of vegetation greening with the corresponding time lag, with a maximum lag of up to four years, as illustrated in Fig. S11a during the 2001-2021 period (Fig. \ref{fig3}b). Approximately 34.3\% of the QTP's total area exhibits significant vegetation growth. Interestingly, about 62\% of these areas experiencing significant vegetation growth are preceded by a precipitation increase at least one year prior. In particular,
in the arid northern regions of the QTP, vegetation exhibits substantial sensitivity to fluctuations in precipitation. In contrast, the southeastern areas, distinguished by plentiful rainfall and verdant vegetation, witness a decrease in vegetation sensitivity in response to an upswing in precipitation (refer to Fig. S11b) \cite{Shen2015GlobChangeBiol}. This differential response can be ascribed to the marginal effect of precipitation. In arid regions, even a minor uptick in rainfall can spur significant vegetation growth. Conversely, in more humid locales, intense storms could inflict damage on the terrain, thereby diminishing vegetation coverage.
Our analysis highlights that the temporal changes in greenness exhibit a significant positive response to precipitation, with a delay of over one year in areas where greenness has experienced significant enhancement ($p<$0.05). In addition to climate change, human afforestation efforts and projects aimed at ecological conservation and restoration, such as the Grain-for-Green Program \cite{delang2016china, Deng2017GEC, Deng2014GCB}, have contributed to land improvement and increased vegetation greenness. Millions of hectares of farmland and degraded land have been converted back to forest and grassland, resulting in an increase of forest ecosystems by $0.6 \times 10^{4} $ km$^{2}$ \cite{li_drivers_2021}.
 Consequently, the mechanism of the greening-induced tipping point in the QTP is likely driven by a combination of climate change and human factors.

Similar to our analysis of the QTP, we examine the Sahel region's vegetation dynamics (Fig. \ref{fig3}g shows the spatial pattern of significant EVI change and Fig. \ref{fig3}h indicates the precipitation-induced greening pattern) over a two-decade span from 2001 to 2021. As for vegetation resilience, Fig. \ref{fig3}i reveals a generally positive trend: the critical EVI$_c$ has increased from 0.17 to 0.19 over the last two decades. This is indicative of a system shifting towards a more resilient tipping state.  However,  it is noteworthy that a minor decline in EVI$_c$ was observed during 2017-2021, likely due to a period of reduced rainfall, aligning with observed climatic patterns. Fig. \ref{fig3}g highlights a macroscopic trend of expanding vegetation cover—often referred to as `greening'—particularly in the  Sahel's northern regions from 2001 to 2021.

Moreover, in Fig. \ref{fig3}j, the empirical value for the fragment size distribution exponent ($\tau$) closely matches theoretical expectations for most of this period, suggesting that the vegetation in the Sahel is operating in a critical state (refer to Figs. S12–S17 for details). Once again, the one noticeable deviation occurs in 2017-2021, which can be attributed to significantly low rainfall levels during that period (refer to Fig. S18). Further corroborating this, the fractal dimensions for mass and external perimeter also align with theoretical values, as shown in Fig. S19. These results support the notion that Sahel's vegetation is near a state of SOC, evidenced by the critical exponents closely matching their theoretical counterparts.

\subsection{Origin of criticality}

To comprehend why the QTP and the Sahel region appear to be on the brink of significant ecological changes, or `tipping points',  we explore two factors: correlation length ($\xi$) and susceptibility ($\chi$), as detailed in Section (\ref{corrxi}). The correlation length measures the extent to which spatial fluctuations in a system are correlated, while susceptibility indicates the system's sensitivity to external influences,  like less rainfall or hotter temperatures. We observe that both $\chi$ and $\xi$ peak at EVI$_c$, as depicted in Figs. \ref{fig3}e and k. This indicates that when the system approaches the critical threshold, it becomes more sensitive and spatially correlated, making it more vulnerable to disturbances.

Furthermore, we find that both $\chi$ and $\xi$ diverge at EVI$_c$ when extrapolated to an infinitely large system size, as illustrated in Figs. \ref{fig3}f and \ref{fig3}l. The divergence of correlation length and susceptibility at EVI$_c$ indicates that the presence of inherent global correlation of the spatio-temporal dynamical processes within the greenness system \cite{Jensen2021JournalofPhysicsComplexity}. This global correlation plays a vital role in the emergence of tipping criticality in the QTP and Sahel. In other words, the increased spatial correlation and sensitivity of the system near the tipping point which is close to EVI$_c$=0.2, lead to stronger responses to external influences, potentially triggering abrupt transitions.

 \subsection{Optimal enhancing resilience model} 
    
 In ecology, maintaining a robust and well-connected giant ecological component (EVI $>$ 0.2) is pivotal as it facilitates essential ecological exchanges such as energy flow, species movement, information transfer, and nutrient cycling. These processes are integral to the health, resilience, and functioning of ecosystems. Energy flow, which initiates from primary producers and traverses various trophic levels, is a fundamental ecosystem process \cite{chapin2002principles}. Similarly, species exchange promotes biodiversity by allowing for migration, dispersal, and colonization \cite{hilty2012corridor}. Information transfer, vital for species adaptation, is enabled through a well-connected ecological network \cite{danchin_public_2004}. Nutrient cycling, crucial for maintaining soil fertility and productivity, also relies on a well-connected ecosystem \cite{chapin2002principles}. 

Hence, protecting the giant ecological component and maintaining its connectivity is essential to preserve the integrity, functionality, and resilience of ecosystems. To this end, we propose an \textit{Optimal Enhancing Resilience Model} (OERM) to strengthen the ecosystem resilience in the QTP and Sahel. This model utilizes the percolation method to identify critical nodes within the system that, when disrupted, can lead to collapse and fragmentation, as shown in Fig. \ref{fig2}. Proactively protecting these nodes is an effective way to reduce systemic vulnerability and risks. In the OERM, we focus on the protection of a single node to minimize socio-economic costs associated with protection efforts. The specific details and methodology for selecting and protecting this single node can be found in Section. \ref{oerm}. 
By focusing on these critical nodes, OERM aims to enhance the resilience and structural integrity of the QTP and Sahel ecosystems, thereby reducing the risk of fragmentation and system failure.

For the QTP during the period 2017-2021, we identify the critical node $i_c$ at the percolation threshold EVI$_c$ that is approximately located at (36.07$^\circ$N, 101.96$^\circ$E), as shown in Fig. \ref{fig4}a and b. We proceed to select $O$ nodes in the critical region and neighboring areas and protect them. 
To quantify the resilience of the QTP, we define the metric $R(O)$,
\begin{equation} \label{EQ_OERM}
R(O) = \frac{G_{1}^{c}(O) - G_{1}^{c}}{G_{1}^{c}},
\end{equation}
where $G_{1}^{c}(O)$ represents the largest cluster size after protecting $O$ nodes, and $G_{1}^{c}$ denotes the largest cluster  pre-protection. Specifically, $G_{1}^{c}(0) = G_{1}^{c}$. Fig. \ref{fig4}c demonstrates a significant resilience boost from $R=0$ to $10.9\%$ for $O \geq 1$, with $O = 1$ emerging as the optimal point. For comparison, we introduced a \textit{Random Enhancing Resilience Model} (RERM), where $O$ nodes are randomly chosen for protection. The results in Fig. \ref{fig4}c (orange curve) reveal that RERM is not effective at improving resilience.


Subsequently, we apply both the OERM and RERM frameworks to the Sahel using data from 2013-2017.
The results, shown in Fig. \ref{fig4}d-f,  confirm that targeted protection of a critical node, as suggested by the OERM, significantly amplifies the system's resilience. Here, protecting just one node ($O=1$),  results in a $R=$23.0\% increase in resilience.



In contrast, RERM does not deliver a comparable boost in resilience, emphasizing the utility of OERM as a pragmatic tool for environmental endeavors. 
This model could guide afforestation efforts by identifying areas that would most benefit from increased vegetation planting, in terms of bolstering overall system resilience. Additionally, it could aid in developing sustainable land use plans by highlighting regions where human activity could potentially push the system towards a negative tipping point. Furthermore, the OERM can assist in prioritizing areas for protection or restoration by singling out those crucial to the ecosystem's overall well-being and robustness.
 
\section{Summary and Discussion}

The present study explored the greening-induced positive tipping point of the Qinghai-Tibetan Plateau (QTP) and Sahel as well as their potential implications.
Our findings unveiled that both the QTP and Sahel are on the verge of a tipping point,  specifically, our calculated percolation threshold EVI$_c$ approximates an empirically determined ecological critical state of roughly 0.2. At this EVI$_c$ threshold, the vegetation architecture adheres to perfect power laws with consistent exponents, encompassing $\tau$, $D_f$, and $D_e$. Importantly, our study discovered that their vegetation undergoes ongoing improvement in system resilience and sustainability, characterizing them as favorable tipping elements. We also propose an Optimal Resilience Enhancement Model. This model aids in boosting vegetation resilience with minimal socio-economic repercussions and presents a practical strategy to preserve the `giant' ecological component. This emphasizes the need for adaptability and transformation in maintaining the resilience of ecosystems \cite{folke2010resilience}.

The greening-induced positive tipping points propose several potential consequences and implications for both the ecosystem and human society. Firstly, in terms of  \textit{ecosystem functioning}, it has the potential to enhance various aspects such as nutrient cycling, water, and energy balance, and habitat availability for wildlife. These alterations contribute to the overall resilience of the ecosystem, enabling it to adapt to future environmental changes \cite{chapin_iii_consequences_2000}. Secondly, it can also influence the provision of essential ecosystem services. For instance, it can impact carbon sequestration \cite{Yun2022JGRB}, water regulation, and soil conservation. Changes in these services have far-reaching implications, particularly for local communities that rely on them for their livelihoods. Furthermore, it can have effects on \textit{climate change feedbacks}. The increased vegetation cover resulting from the tipping points can bring about alterations in the regional climate by influencing surface albedo, evapotranspiration, and the overall energy balance \cite{bonan_forests_2008}. These changes can have cascading effects on global climate patterns, potentially mitigating risks induced by climatic changes.

Despite these promising findings, concerns persist regarding the future prospects of the QTP and Sahel vegetation due to potential changes in climate conditions and continued warming \cite{Ehlers2022ER}. For example, the ongoing loss of glacier water in QTP is particularly worrisome as it may result in water shortage in the future, which could disrupt the stability of vegetation growth \cite{Immerzeel2019Nature}. Additionally, the increasing frequency of vegetation fires poses a significant threat to ecological stability, with young trees being particularly vulnerable to high-severity fires attributed to persistent warming \cite{bowman_vegetation_2020}. Notably, fire regimes are not only influenced by climate change but could also lead to net carbon emissions from terrestrial carbon reservoirs.

Our findings shed light on the positive tipping phenomenon in vegetation resilience amidst climate change and human impact, emphasizing the urgent need for robust strategies to strengthen ecosystem resilience and promote sustainable development. Employing a percolation-based framework and resilience analytics, our study provides guidance for policymakers and environmental managers in preserving ecosystems and protecting the socioeconomic welfare of future generations.



\clearpage

\begin{figure}
\begin{centering}
\includegraphics[width=0.9\linewidth]{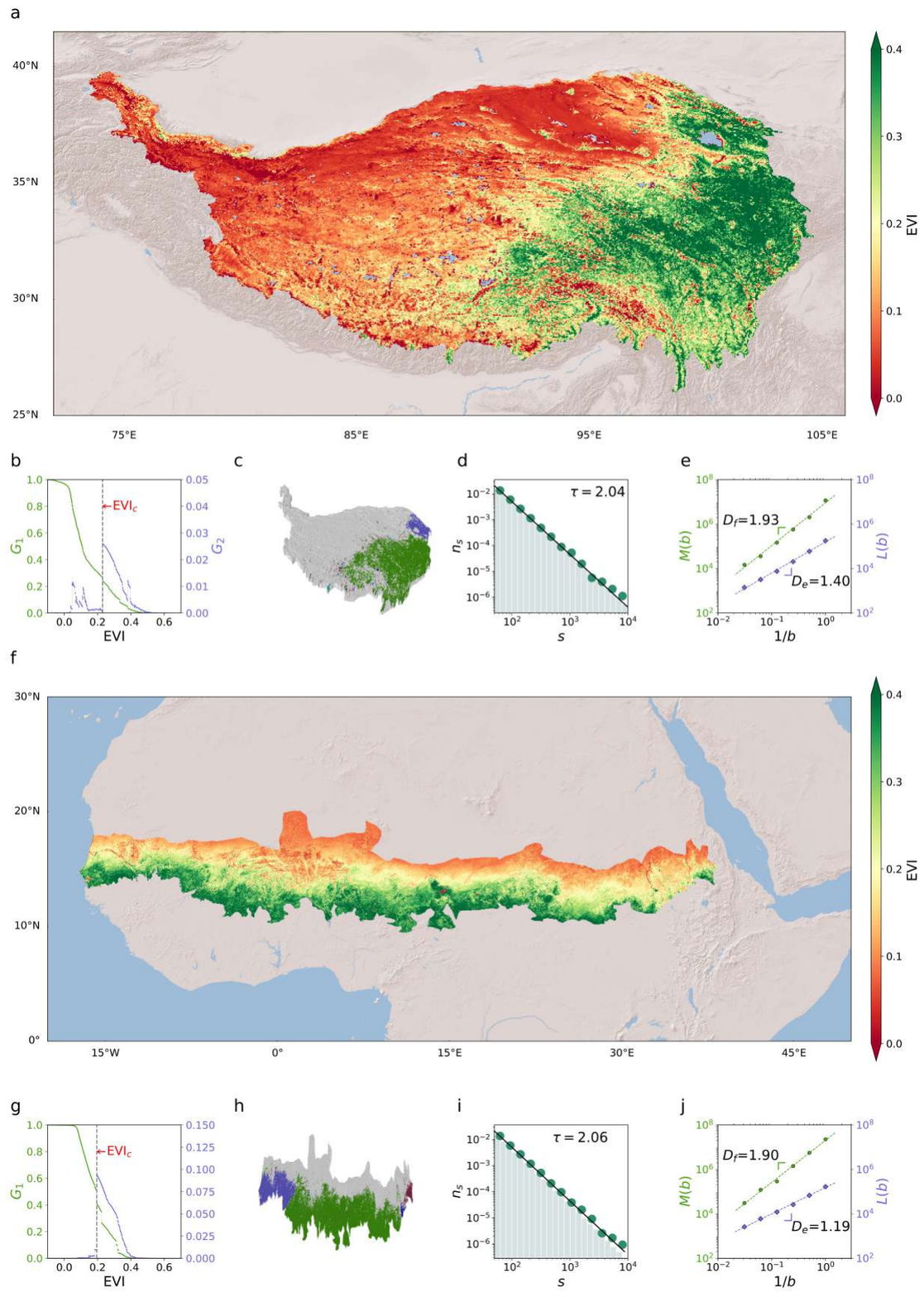}
\caption{\label{fig2} 
{\bf Criticality of vegetation in the Qinghai-Tibetan Plateau and the Sahel.
}}
\end{centering}
\end{figure}
\begin{figure}[t]
  \contcaption{{\bf a},   
Spatial distribution of the average EVI across the QTP over 2017-2021. EVI is an indicator of an area's greenness, with higher values indicating denser, healthier vegetation.  {\bf b}, Evolution of the relative sizes of the largest (denoted by $G_1$, in green) and second largest (denoted by $G_2$, in blue) fragments in relation to EVI. The vertical dashed line indicates the percolation threshold, EVI$_c$. {\bf c}, Fragmentation patterns of greenness at EVI$_c$. Different colours depict distinct fragments, with fragments of over 100 nodes displayed. The largest fragment is illustrated in green and the second largest in blue.
{\bf d}, Observed (green dots) and null model (gray histogram) distribution of fragment sizes at EVI$_c$. The solid line represents the power-law fit with an exponent $\tau$ $\approx$ 2.04 ($R^2$=0.996). Statistics incorporate approximately 87,000 fragments.
{\bf e}, Mass ($M$, green dots) and external perimeter ($L$, blue dots) for the largest fragment at various rescaled box renormalization factors ($b=2^0,\ 2^1,\ 2^2,\ 2^3,\ 2^4,\ 2^5$ ). Power-law fits with exponents $D_f\approx 1.93$ ($R^2$ = 0.904) and $D_e \approx 1.40$ ($R^2$ = 0.971) are denoted by the green and blue dashed lines, respectively.
{\bf f}-{\bf j}, The same as {\bf a}-{\bf e}, but for the Sahel region over 2013-2017.}
\end{figure}

\begin{figure}
\begin{centering}
\includegraphics[width=1.0\linewidth]{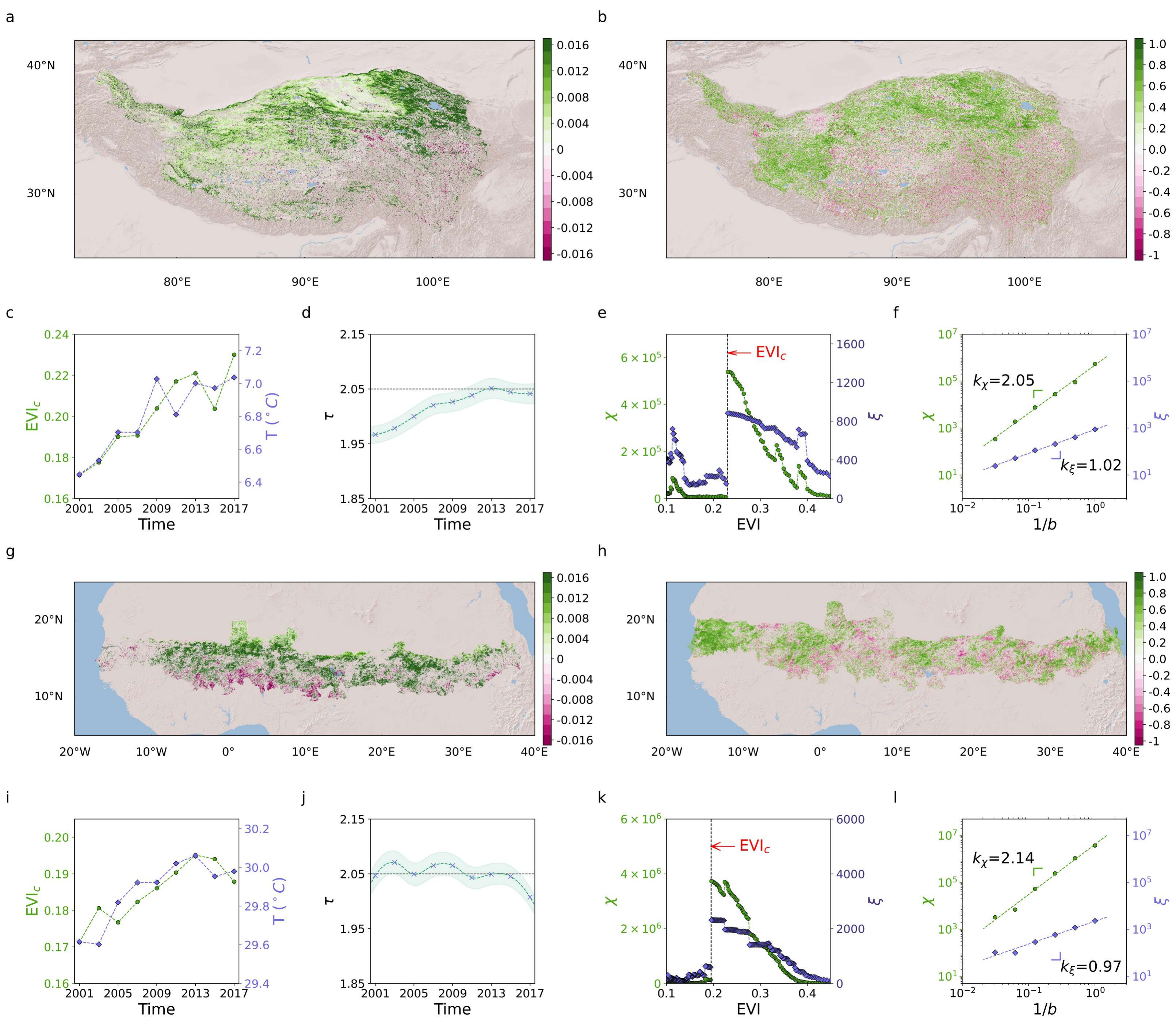}
\caption{\label{fig3} {\bf Greening-induced tipping point in the Qinghai-Tibetan Plateau and the Sahel.} }
\end{centering}
\end{figure}
\begin{figure}[t]
  \contcaption{{\bf a}, The spatial pattern of significant EVI change trends ($p<0.05$) from 2001 to 2021, demonstrating areas of remarkable vegetation change.  {\bf b}, Precipitation-induced greening patterns from 2001 to 2021. The cross-correlation coefficients (according to Eq. (\ref{redefine_CC})) between precipitation and EVI series are computed, revealing regions with significant correlation ($p<0.05$). The colours denote the corresponding cross-correlation coefficients. EVI series in the northern arid regions of the QTP display high sensitivity to precipitation variations. However, in the southeast regions characterized by high precipitation and lush vegetation, increased precipitation is linked to decreased vegetation sensitivity (refer to Fig. S11b). {\bf c}, A comparison of EVI$_c$ (green dots) and the average temperature $T$ in June-August (blue dots) as functions of time. {\bf d}, The temporal evolution of the critical exponent $\tau$ over the past two decades (2001-2021). The green shade represents the error bars. The horizontal black dashed line indicates the theoretical value of $\tau \approx 2.05$ as predicted by classical 2D percolation theory. Each point is calculated within a sliding window of 5 years.  {\bf e}, The  correlation length $\xi$ (blue dots) and  susceptibility $\chi$ (green dots) with respect to EVI. The vertical black dashed line indicates EVI$_c$.
  {\bf f}, The behavior of $\xi$ (blue dots) and $\chi$ (green squares) as functions of $1/b$, where $b$ represents the rescaled box renormalization factor. The best-fit lines for $\xi$  and $\chi$  exhibit slopes of $k_\xi=1.02$ ($R^2=0.998$) and $k_\chi=2.05$ ($R^2=0.901$), respectively.
Panels {\bf g}-{\bf l}, The same as {\bf a}-{\bf f}, specifically for the Sahel region.}
\end{figure}

       \begin{figure}[ht]
          \centering
          \includegraphics[width=1.0\linewidth]{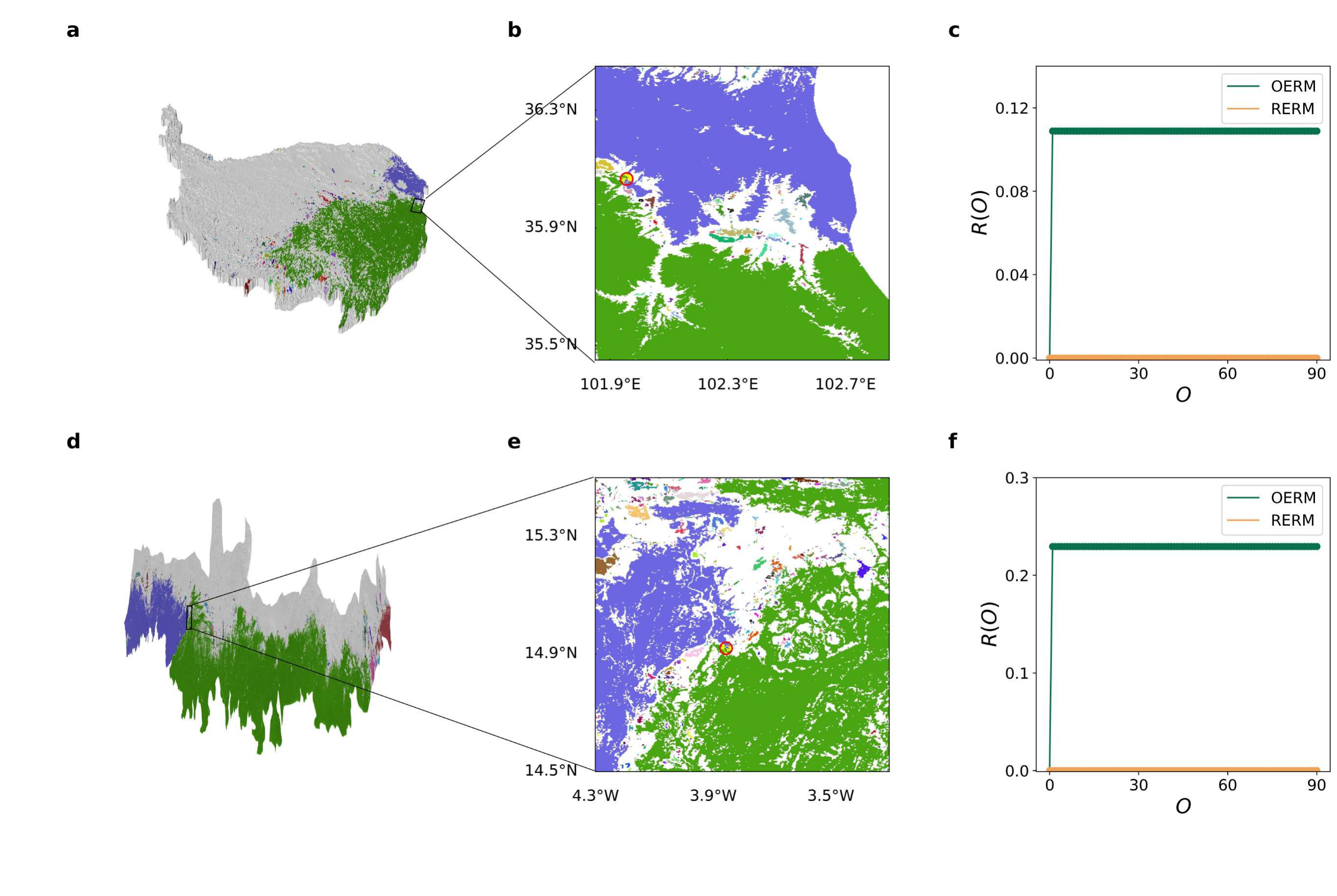}
          \caption{ {\bf Optimal Enhancing Resilience Model}.
{\bf a,} Spatial fragmentation patterns of the QTP at the critical node identified using EVI$_c$ for the period 2017-2021. 
The vulnerable area is highlighted by a black rectangle. {\bf b,}
Location of the critical node $i_c$ in the QTP. The node, situated at  (36.07$^\circ$N, 101.96$^\circ$E), is indicated by a red circle within the black rectangle in panel {\bf a}. {\bf c,} The Resilience $R$, as quantified by Eq. (\ref{EQ_OERM}), is depicted as a function of protection nodes $O$, for both OERM (represented by the green line) and the Random Enhancing Resilience Model (indicated by the yellow line) for the QTP during 2017-2021. Resilience increases from 0 to $10.9\%$ when $O=1$, indicating the effectiveness of the OERM in enhancing system resilience. {\bf d-f,} Analogous to panels {\bf a-c} for the QTP, but for the Sahel region during the period 2013-2017.
The critical node $i_c$ is located at (14.92$^\circ$N, 3.86$^\circ$W) and resilience increases from 0 to $23.0\%$ when $O=1$. }
          \label{fig4}
        \end{figure}

\clearpage
    \section{Data and Methods}
    \subsection{Data} \label{datasection}
In this study, we utilize the Enhanced Vegetation Index (EVI) data derived from Moderate Resolution Imaging Spectroradiometer (MODIS) satellite imagery. EVI is a remote sensing metric that quantifies the greenness or density of vegetation on the Earth's surface. It considers the reflectance of visible and near-infrared light to optimize the vegetation signal while minimizing atmospheric and background noise. Compared to other vegetation indices like the Normalized Difference Vegetation Index (NDVI), EVI is more effective in areas with dense vegetation or under varying atmospheric conditions. EVI values range from -1 to 1, with higher values indicating healthier and more vigorous vegetation \cite{Huete2002RemoteSensingofEnvironment}. We obtained the spatial composites provided at 16-day temporal resolution and 250-meter spatial resolution (MOD13C1 Version 6; \url{https://lpdaac.usgs.gov/products/mod13q1v006/}).

Additionally, we incorporate monthly surface air temperature reanalysis data ($0.1^\circ\times 0.1^\circ$) from ERA5 Land (\url{https://doi.org/10.24381/cds.68d2bb30}) and monthly precipitation accumulation reanalysis data ($0.04^\circ\times 0.04^\circ$) from TerraClimate \cite{Abatzoglou2018} (\url{https://www.climatologylab.org/terraclimate.html}).

These data were preprocessed and downloaded on the Google Earth Engine (GEE) cloud platform  \cite{gorelick2017google}. And, we focus on the Qinghai-Tibetan Plateau (QTP) area (\url{https://doi.org/10.3974/geodb.2014.01.12.V1}, accessed on 31 July 2022) and Sahel area.
The Sahel region selects areas with EVI ranging between 0.1 and 0.4, corresponding to the shapefile provided in the attachment.
We took summer (July-September) as the focused season. The summer months represent the peak growing season on the QTP/Sahel, making it crucial for studying vegetation patterns, and observing the highest proportion of vegetation cover in the region.

\subsection{Data filtering}

To minimize the effects of annual fluctuations in our study, we employ a sliding-window average approach. We use a 5-year sliding window that progresses by 2 years at a time, resulting in adjacent windows having a temporal overlap of 3 years. This approach calculates the temporal average of EVI values pixel-by-pixel within each sliding window. By smoothing out short-term variations and noise, the sliding-window average provides a more stable and reliable representation of the underlying spatial patterns and trends in the data. The overlapping windows ensure a continuous and smooth analysis of the temporal changes, minimizing the potential for abrupt shifts or artificial breakpoints in the time series. Each window is defined by the start year of the corresponding period.

\subsection{Greenness Percolation Model}\label{PercolationOperation} 

In our Greenness Percolation model, we represent the EVI image as a 2D lattice network, where each \textit{valid} pixel in the image corresponds to a node in the network. Valid pixels are those located within the boundary of the QTP (Sahel) and have non-null EVI values. Each image contains approximately 48.5 (55.2) million valid pixels, providing a high enough spatial resolution suitable for percolation analysis.

The percolation procedure follows these steps:
        starting from a fully occupied lattice network, we rank all the nodes based on their EVI values in ascending order. We then remove nodes one by one, starting with the node having the lowest EVI value and progressing to higher-ranked nodes. During the removal process, we identify the fragments using classical percolation theory  \cite{Cohen2010ComplexNetworks,eaNewman2018ComplexNetwork}. A fragment is defined as a subset of nearest neighbour nodes, where each node in the subset is connected to at least one other node.
        We employ the efficient Newman–Ziff algorithm  \cite{Newman2000PhysRevLett} to detect the different fragments at each step.
        
It's worth mentioning that our model incorporates free boundary conditions. Each node interacts with four neighbors, excluding the nodes adjacent to invalid pixels or the boundaries of the QTP/Sahel. In percolation theory,  the relative size of the largest fragment, often referred to as the `giant' fragment, is typically identified as the order parameter  \cite{Cohen2010ComplexNetworks}.
Here, the `giant' fragment $G_1$ is defined as  \cite{Fan2018ProceedingsoftheNationalAcademyofSciences},
\begin{equation}
            G_1 = \frac{\max \left [ 
                            \left ( \sum_{i \in \mathcal{H}_1} g_i \right ),  
                            \left ( \sum_{i \in \mathcal{H}_2} g_i \right ),    
                            ...,
                            \left ( \sum_{i \in \mathcal{H}_m} g_i \right ),  
                            ...,
                            \right ]}{\sum_{i=1}^{N} g_i}
                            ,
\label{giant_fragment}
\end{equation}
where $\mathcal{H}_m$ denotes a series of disjoint fragments, $g_i$ stands for the size/area of node $i$, and $N$ is the total number of nodes.
Since most of all nodes in the QTP are located within $[30^\circ N, 40^\circ N]$, and most of all nodes in the Sahel region are located within $[10^\circ N, 20^\circ N]$, We assume their sizes are approximately equal, i.e., $g_i = 1$; However, if the latitude spans larger scales, we use then $g_i = \cos{(\phi_i})$, where $\phi_i$ is the latitude of node $i$ \cite{Fan2018ProceedingsoftheNationalAcademyofSciences}. We have also performed the same analysis by using $\phi_i$, and the results do not affect our conclusion. Throughout the percolation simulation, we measure the order parameter $G_1(P)$ at each time step $P$ and calculate the largest one-step gap $\Delta_c$ as defined by Equation (\ref{gapc}),
        \begin{equation}\label{gapc}
            \Delta_c \equiv \max_{P} \left [G_1(P-1) - G_1(P) \right ].
        \end{equation}
Equation (\ref{gapc}) is used to determine the percolation threshold \cite{fan_universal_2020}.        
We define the time step with the largest jump $\Delta_c$ as $P_c$. 
The EVI value of the critical node at $P_c$ is defined as the percolation threshold EVI$_c$, at which the greenness state abruptly shifts from cohesiveness to fragmentation.

\subsection{Power-law fitting for the fragment-size distribution}\label{clustersizefitting}
In our analysis, we calculated the fragment-size distribution at the EVI$_c$. We used the maximum-likelihood fitting method \cite{Clauset2009SIAMReview} to fit the frequency of fragment sizes to a power-law distribution according to Eq. (\ref{EQ4}). The fitting was performed using a Python package  \cite{Alstott2014PLOSONE} available online,
        \begin{equation}\label{EQ4}
            n_s\sim s^{-\tau}
        \end{equation}
where $n_s$ is the number of fragments of size $s$.
We assess the Goodness of fit using the Kolmogorov-Smirnov statistic and likelihood ratios.
In the main body of our paper, we set the maximum size of the power-law fitted distributions $s_{\mathrm{max}}$ to 10,000. Additionally, we test alternative maximum sizes, 5,000, 6,000, 7,000, 8,000, 9,000, and 11,000 for the QTP and 9000, 11,000, 12,000, 13,000, 14000, and 15,000 for the Sahel. The results show robustness and consistency, as demonstrated in Figs. S20 and S21, supporting the findings and conclusions of the study. These consistent results provide confidence in the methodology used in the research.

    \subsection{Box renormalization}\label{renorm}

Box renormalization approach was employed to quantify the self-similarity features of the largest fragment at the percolation threshold \cite{stauffer2018introduction}.
This procedure involves systematically reducing the resolution of a large lattice grid by eliminating fluctuations on scales smaller than a given length scale, $b$. The goal is to re-scale the lattice grid into a sequence of smaller ones, enabling the study of properties such as fractal behaviour and finite-size effects. The box renormalization operation involves selecting a box size $b$ (e.g., $b=2^1,\ 2^2,\ 2^3,\ 2^4,\ 2^5$, ...), and re-grouping the lattice grid into $b \times b$ boxes. Each box is replaced with a super-node, where the super-node's value (EVI) is the average of all the nodes in the corresponding box.
By applying box renormalization with a sequence of re-scale factors $b$, a sequence of re-scaled lattice grids is obtained. The percolation procedure can then be re-applied to these re-scaled lattice grids to examine properties such as fractal behaviour and finite-size effects. For consistency, the re-scale factor $b$ for the original lattice grid is defined as 1.

\subsection{Fractal dimensions}\label{fractaldimension}
The fractal concept, first introduced by Mandelbrot \cite{MandelbrotFractal}, was applied to percolation by Stanley  \cite{Stanley1977JournalofPhysicsAMathematicalandGeneral} to describe the complexity of fractal substructures within the giant fragment at the percolation threshold. At this threshold, the giant fragment exhibits self-similarity on all length scales and can be considered a fractal. Here, we focus on two properties: the mass and external perimeter of the giant fragment at the percolation threshold.

The mass of the largest fragment is defined as the number of nodes it contains, while the accessible external perimeter, introduced by Aharony \cite{Grossman1987JournalofPhysicsAMathematicalandGeneral}, is defined as the number of nodes on the external boundary. The external boundary can be determined by probing (adsorbent) nodes that are adjacent to the largest fragment’s exterior.

To estimate the mass and external perimeter fractal dimensions ($D_f$ and $D_e$, respectively), we examine the power-law relationship between these quantities and the re-scale factors $b$ as given by Eqs. (\ref{df}) and (\ref{de}) where the mass and external perimeter of the giant fragment at the percolation threshold are represented by $M(b)$ and $L(b)$ respectively,
\begin{equation}\label{df}
M(b) \sim b^{-D_f},
\end{equation}
\begin{equation}\label{de}
L(b) \sim b^{-D_e}.
\end{equation}
The values $D_f$ and $D_e$ indicate the presence of self-similar fractal patterns and provide insight into the complexity of the fractal substructures within the fragment.

To obtain the mass and external perimeter of the giant fragment at EVI$_c$, we renormalize the EVI image for a sequence of re-scale factors ($b=2^1,\ 2^2,\ 2^3,\ 2^4,\ 2^5$). The EVI$_c$ value is defined as the EVI corresponding to the largest jump of the giant fragment within the selected EVI range of 0.15-0.3, which is mainly characterized by the alpine meadow. Finally, we calculate the mass and external perimeter fractal dimensions of the largest fragment at EVI$_c$ using Eqs. (\ref{df}) and (\ref{de}) for $b=2^0,\ 2^1,\ 2^2,\ 2^3,\ 2^4,\ 2^5$.   

    \subsection{Null model}\label{nullmodel}
In this study, we create a \textit{null model} to serve as a comparison standard. The null model is generated by randomly shuffling the EVI values to create 100 new samples. The percolation procedure was then applied to each of these new samples. In these new samples, the original spatial associations are destroyed, and percolation is performed on a random spatial profile, corresponding to the uncorrelated site percolation class.
According to percolation theory, the null model is expected to exhibit critical phenomena at the percolation threshold. Therefore, it serves as a comparison standard for critical features. By analysing the behaviour of the null model, we can better understand the significance of spatial associations in the original data and assess their impact on the observed critical phenomena and tipping points in the QTP and Sahel ecosystems.
        
    \subsection{Correlation length and susceptibility}\label{corrxi}
The correlation length ($\xi$) in percolation theory represents the average distance between two sites that belong to the same fragment. Equation (\ref{xi-define}) defines the correlation length as,
\begin{equation}{\label{xi-define}}
\xi^2 = \frac{2\sum'_s R_s^2 s^2 n_s}{\sum'_s s^2 n_s},
\end{equation}
where $n_s$ denotes the number of fragments of size $s$, $R_s$ is the average gyration radius of fragments of size $s$, and the prime on the sums indicates the exclusion of the largest fragment in each measurement.

The gyration radius of a given fragment $a$ of size $s$ is defined by,
\begin{equation}
R^2_a = \frac{1}{s^2} \sum_{i=1}^{s}\sum_{j>i}^s |\mathbf{r}^a_i - \mathbf{r}^a_j|^2,
\end{equation}
The average gyration radius of all the fragments of size $s$ can be calculated using Equation (\ref{radius-define}),
\begin{equation}{\label{radius-define}}
R_s^2 = \frac{1}{s^2 n_s} \sum_{a=1}^{n_s}\sum_{i=1}^{s} \sum_{j>i}^{s}|\mathbf{r}_i^a - \mathbf{r}_j^a|^2,
\end{equation}
where $|\mathbf{r}_i^a - \mathbf{r}_j^a|$ denotes the Euclidean distance between the $i$th and $j$th node in fragment $a$.
The susceptibility $\chi$ is defined as:
\begin{equation}
\chi = \frac{\sum'_s s^2 n_s}{ \sum'_s s n_s},
\end{equation}
where the prime on the sums indicates the exclusion of the largest fragment in each measurement. The susceptibility represents the system's sensitivity to changes in the percolation threshold, and it is used to characterize the critical phenomena.

To demonstrate that the correlation length $\xi$ and susceptibility $\chi$  diverge at EVI$_c$ as the size of the empirical system tends to approach infinity, we investigate the finite-size effects using box renormalization. We renormalized the EVI image for a sequence of re-scale factors $b=2^1,\ 2^2,\ 2^3,\ 2^4,\ 2^5$ and applied the percolation procedure to these re-scaled EVI images. Then we calculated the correlation length $\xi$ and susceptibility $\chi$ at EVI$_c$ for $b=2^0,\ 2^1,\ 2^2,\ 2^3,\ 2^4,\ 2^5$, examining their behaviour as the system size is extrapolated to infinity.
By analysing the behaviour of $\xi$ and $\chi$ as a function of the re-scale factor $b$, we can determine whether these quantities diverge at the critical EVI value (EVI$_c$), indicating the existence of a critical point and a phase transition at the percolation threshold.

\subsection{Cross-correlation}
The cross-correlation with time lag $\sigma$ between the EVI series $E(y)$ and the precipitation series $Q(y)$ are defined by:
        \begin{equation}
            C(\sigma) = \frac{\langle Q(y) E(y+\sigma) \rangle - \langle Q(y)\rangle \langle E(y+\sigma) \rangle}{\sqrt{\left \langle \left (  Q(y)- \langle Q(y)\rangle \right)^2 \right \rangle}\times\sqrt{\left \langle \left (  E(y+\sigma)- \langle E(y+\sigma)\rangle \right)^2 \right \rangle}}
        \end{equation}
where $\sigma \in [-\sigma_\mathrm{max}, \sigma_\mathrm{max}]$ is the time lag, with $\sigma_\mathrm{max}=4 $ year.
        Therefore, we can achieve $2\sigma_\mathrm{max}+1$ different cross-correlation values.
The maximum absolute value of the cross-correlation function was identified, and the corresponding time lag is denoted as $\sigma_0$. 
In this study, we redefine cross-correlation $C$ as, 
\begin{equation}
C = C(\sigma_0).
\label{redefine_CC}
\end{equation}

        \subsection{Optimal enhancing resilience model}\label{oerm}
We propose the Optimal Enhancing Resilience Model (OERM) as a solution to further improve the stability of the system. The OERM is based on the snapshot of fragments at EVI$_c$ where the critical node $i_c$, is identified as the node connecting the largest and second-largest fragments at EVI$_c$. The total number of protected nodes is denoted by $O$ (with a max value of $90$ in this study). When the protection steps $O=0$, no nodes are specifically protected. When the protection steps $O=1$, the critical node $i_c$ is protected, and its EVI value is changed to $1.0$. Subsequently, unoccupied nodes at EVI$_c$ are gradually protected based on their distance from the critical node $i_c$, following a near-to-far approach. If multiple nodes have the same distance, the one with a larger EVI is protected first. The EVI of the protected nodes is also changed to $1.0$.

The protection resilience $R$, is defined as the ratio of the fragment size increase after protection to the fragment size before protection. It is measured using Eq. (\ref{EQ_OERM}). To compare the effectiveness of OERM, we propose the Random Enhancing Resilience Model (RERM), where an equal number of protected nodes are randomly selected.
By comparing the performance of OERM and RERM, we aim to evaluate the effectiveness of targeted protection strategies based on critical nodes and their surrounding areas. This comparison would help develop more efficient and cost-effective measures for enhancing ecosystem resilience and mitigating the impacts of climate change.

\section*{Data availability}
The data used in Figs. 2–4 are available as Source Data, and other data supporting the plots within the paper and findings of the study can be obtained from the corresponding author upon request.

\section*{code availability}
The Python codes used for the analysis are available on GitHub (\url{https://github.com/fanjingfang/RGPTP}).

\section*{Acknowledgments}
We acknowledge the support of the National Natural Science Foundation of China (Grant No. 12275020, 12135003, 12205025) and the Ministry of Science and Technology of China (2019QZKK0906). D.C. is supported by the Swedish strategic research area MERGE. J.K. received support from the Germany BMBF grant 01LP1902A.

{\section*{Author Contributions}
J.F. designed the research. Y.S. and J.F. performed the analysis and prepared the manuscript, Y.S., T.L., S.W., J.M., Y.Z., S.Y., X.C., D.C., J.K., S.H., H.J.S. and J.F. discussed results and contributed to writing the manuscript. J.F. led the writing of the manuscript. }

\section*{Additional information}
Supplementary Information is available in the online version of the paper.

\section*{Competing interests}
The authors declare no competing interests.

\clearpage
\bibliographystyle{naturemag}
\bibliography{Main_V4}
\end{document}